\documentclass[useAMS,usenatbib]{mn2e} \usepackage{amsmath}
\usepackage{amssymb} \usepackage{graphicx}

\newcommand\bb[1] { \mbox{\boldmath{$#1$}} }
  
 \newcommand{\beq}{
\begin{equation} } \newcommand{\eeq}{ \end{equation} }

\def\spose#1{\hbox to 0pt{#1\hss}}
\def\ltsim{\mathrel{\spose{\lower.5ex\hbox{$\mathchar"218$}}
\raise.4ex\hbox{$\mathchar"13C$}}}

\title[Sedimentation of large solid bodies in protoplanetary discs]
{Midplane sedimentation of large solid bodies in turbulent
protoplanetary discs}

\author[Carballido, Fromang \& Papaloizou] {Augusto Carballido$^1$,
S\'ebastien Fromang$^{2}$ and John Papaloizou$^{2}$ \\ $^1$ Institute
of Astronomy, University of Cambridge, Madingley Road, Cambridge, CB3
0HA, UK \\ $^2$ DAMTP, University of Cambridge, Centre for
Mathematical Sciences, Wilberforce Road, Cambridge, CB3 0WA, UK }

\date{Accepted.  Received; in original form }

\pubyear{}

\begin{document} 

\maketitle

\begin{abstract} 
We study the vertical settling of solid bodies in a turbulent
protoplanetary disc. We consider the situation when the coupling to
the gas is weak or equivalently when the particle stopping time
$\tau_{st}$ due to friction with the gas is long compared to the
orbital timescale $\Omega^{-1}$.  An analytical model, which takes
into account the stochastic nature of the sedimentation process using
a Fokker-Planck equation for the particle distribution function in
phase space, is used to obtain the vertical scale height of the solid
layer as a function of the vertical component of the turbulent gas
velocity correlation function and the particle stopping time. This is
found to be of the same form as the relation obtained for strongly
coupled particles in previous work.

We compare the predictions of this model with results obtained from
local shearing box MHD simulations of solid particles embedded in a
vertically stratified disc in which there is turbulence driven by the
MRI. We find that the ratio of the dust disc thickness to the gas disc
thickness satifies $H_d/H=0.08 (\Omega \tau_{st})^{-1/2}$, which is in
very good agreement with the analytical model. By discussing the
conditions for gravitational instability in the outer regions of
protoplanetary discs in which there is a similar level of turbulence,
we find that bodies in the size range $50$ to $600$ metres can
  aggregate to form Kuiper belt--like objects with characteristic
  radii ranging from tens to hundreds of kilometres.
\end{abstract}


\section{Introduction}

Dust is the constituent of planetesimals, that are believed to lead to
the formation of planets in our own and other solar
systems. Observational evidence for growth of dust particles in
protoplanetary discs comes from near and mid-infrared imaging,
mid-infrared spectrometry and millimetre interferometry
\citep{nattaetal06,rodmannetal06}. They point to the presence of
grains of millimetre and even centimeter size in discs around clasical
T Tauri stars.

In laminar discs, it is well known that gas drag causes the particles
to sediment towards the disc midplane. In the absence of turbulence,
there is nothing to oppose sedimentation and a very thin dust
sub--layer can form. Thus there is a possibility that gravitational
instability ocurs giving rise to gravitationally bound clumps,
eventually leading to planetesimal formation
\citep{goldreich&ward73}. But collective effects arising through the
interaction of the optically thick sedimenting dust layer with the gas
may lead to the generation of turbulence and inhibit gravitational
instability \citep[see for
example][]{weidenschilling80,cuzzietal93,youdin&shu02,gomez&ostriker05}.

On the other hand, provided an adequate degree of ionisation can be
maintained, accretion flows in which the dominant motion is Keplerian
rotation, such as occur in protoplanetary discs, are not laminar. They
have been shown to develop turbulence as a result of the
magnetorotational instability \citep[MRI;][]{hawleyetal95}. The
gaseous velocity fluctuations associated with the turbulence affect
the spatial distribution of particles. Small particles are strongly
coupled to the fluid and essentially follow the gas.

The radial diffusion of a passive contaminant was studied by
\citet{carballidoetal05}, who found a diffusion coefficient
approximately 10 times smaller than the effective disc kinematic
viscosity associated with angular momentum transport.  As shown
recently by \citet{johansenetal06}, this ratio, found to be of order
unity in other simulations \citep{johansen&klahr05}, is dependent on
the topology of the initial magnetic field and the amount of conserved
magnetic flux present, which determines the level of the resulting
turbulence.  They note that the form of the particle diffusion induced
by the turbulence is apparently determined by this level, or
equivalently, the amount of angular momentum transport.
\citet{turneretal06} also studied the vertical spreading of a trace
species in the upper layers of a stratified disc. They found radial
spreading to be faster than vertical spreading throughout the vertical
extent of the disc.

 Larger particles, for which the coupling is weaker, are found to
start to sediment despite the presence of turbulence. For a surface
density $\Sigma = 200$ g cm$^{-2}$ being characteristic of the minimum
mass solar nebula at $5.2$ AU, sedimentation was found to begin for
particle sizes between a centimetre and a metre in a disc model with
MRI driven turbulence \citep[][hereafter FP]{fromang&pap06}. FP
modelled this settling process by means of an advection-diffusion
equation for the dust density. They expressed the diffusion
coefficient in terms of the turbulent gas velocity correlation
function. They found that turbulence stirs the dust sub-layer
significantly, up to approximately 20\% of the disc scale height for
particles of about 10 cm in radius at $5.2$ AU. Thus dust particles of
this size range could not be the constituents of a gravitationally
unstable sub-disc. Accordingly, on a scale of $5$ AU in a minimum mass
solar nebula, the formation of planetesimals is hindered until a much
larger particle size range is reached when the MRI operates.

The above studies modelled the dust component as a second
fluid. However, to study the effect of MHD turbulence on solids of
larger size such that the stopping time due to gas drag significantly
exceeds the inverse orbital angular frequency, it is necessary to
model dust grains as discrete particles.  Johansen et al. (2006) have
performed local three-dimensional simulations of a non-stratified
turbulent protoplanetary disc and find that individual particles tend
to concentrate by a factor of up to 100 due to radial concentration in
vortices.  They suggest that such dust agglomerations can potentially
become gravitationally unstable. However, their study did not address
the effect of the vertical sedimentation of solid bodies.

It is the purpose of this paper to perform a study of the stochastic
vertical settling of solids that is applicable when the stopping time
is much larger than the orbital period.  Firstly we develop an
analytical model based on solving a one dimensional Fokker--Planck
equation. The derivation of this equation and its solution are
presented in Section~2. This gives an expression for the dust
sub--disc thickness as a function of particle size. In Section~3, we
go on to compare these results, and others found in previous work
applicable to strongly coupled small particles, with those found from
simulations of solid particles embedded in a turbulent protoplanetary
disc. Simulations have been performed which treat the dust dynamics
using the Epstein and Stokes drag laws \citep{weidenschilling77}. Note
that these include locally induced radial motion, but not that induced
by a global radial pressure gardient, so that in the former context,
the vertical settling that is found will occur regardless of radial
concentration in vortices as long as the particles do not
collide. Finally, in Section 4 we discuss the implications of our
results for the formation of planetesimals and the possibility of
gravitational instability in the outer regions of protoplanetary
discs.

\section{The model}
\label{model}

In a protoplanetary disc, solid bodies evolve under the combined
influence of gas drag and MHD turbulence. By damping motion relative
to the gas, on average the former results in settling towards the
equatorial plane of the disc. Simultaneously, the latter excites
random motions that oppose the settling process.  The ensemble of
solid bodies will eventually reach a steady state with an associated
vertical semi-thickness $H_d$. Below we present a model to determine
this equilibrium state which is applicable when the coupling of the
particle to the gas is weak and such that the stopping time is very
much longer than the orbital period.

\subsection{Definitions and notation}

The interaction between the gas and the solid bodies (dust, boulders,
planetesimals) occurs through a drag force $\bb{F_D}$. This can be
written as
\begin{equation}
\bb{F_D}=m_p \frac{\bb{u}-\bb{v}}{\tau_{\rm{st}}} \, ,
\end{equation}
where $m_p$ is the mass of a particle, $\bb{v}$ its velocity and
$\bb{u}$ is the gas velocity.

The characteristic time required to equilibrate gas and particle
velocities, or stopping time is $\tau_{\rm{st}}.$ This depends on the
ratio of the size of the particle, expressed through the radius $a$
they would have if they were spherical, and the mean free path of the
gas molecules, $\lambda$ \citep{weidenschilling77,cuzzietal93}.  When
$\lambda>(4/9)a$, we are in the Epstein regime for which
\begin{equation}
\tau^{(Ep)}_{\rm{st}}=\frac{\rho_s}{\rho} \frac{a}{c_s} \, ,
\label{epstein_tau}
\end{equation}
where $\rho_s$ is the solid mass density and $c_s$ is the gas speed of
sound.

 When $\lambda<(4/9)a$, we are in the Stokes regime, and
 $\tau^{(St)}_{\rm{st}}$ is defined through
\begin{equation}
\tau^{(St)}_{\rm{st}}=\frac{8}{3}\frac{\rho_s}{\rho}\frac{a}{C_D
|\bb{u}-\bb{v}|} \, ,
\label{stokes_tau}
\end{equation}
where $C_D$ is a dimensionless coefficient that depends on the
Reynolds number of the flow ${\cal R}_e,$ tending to a constant value
of order unity for ${\cal R}_e \rightarrow \infty.$

\subsection{Diffusion in velocity--space}
\label{diffusion vel}

Since we are interested in the problem of vertical settling, we
 restrict the analysis to consider only the vertical direction $z,$
 though we shall consider relevant results of simulations
 relating to the horizontal directions below.
  
  The vertical component of the equation of motion for a single
particle moving under the gravitational acceleration due to the
central star and gas drag can be written as
\begin{equation}
\frac{dv_{z}}{dt}=-\Omega^{2}z+\frac{u_{z}-v_{z}}{\tau_{\rm{st}}}
\label{eq:fp1}
\end{equation}
where $u_{z}$ is the vertical component of the gas velocity which is
fluctuating stochastically because it is turbulent. Below we make the
approximation that $\tau_{\rm{st}}$ is constant. In reality, in the
Epstein regime, density fluctuations will cause
$\tau^{(Ep)}_{\rm{st}}$ to vary with time. In the Stokes regime, both
density and velocity fluctuations will have the same effect. However,
we expect the essential physics to be retained if we adopt an
appropriate time averaged value of $\tau_{\rm{st}},$ so we assume this
to have been done in what follows.
 
Equation~(\ref{eq:fp1}) can be written as an equation for a damped
 harmonic oscillator randomly forced in time by the part of the drag
 force proportional to the gaseous turbulent velocity field:
\begin{equation}
\frac{dv_z}{dt}+\frac{v_z}{\tau_{\rm{st}}} + \Omega^2
z=\frac{u_{z}}{\tau_{\rm{st}}}\equiv F \, .
\label{eq:fp2}
\end{equation}
The particle motion can be regarded as driven by the stochasting
forcing term $F$. When the correlation time of the gas velocity is
much smaller than the orbital period, this can be regarded as
producing a series of weak (which will be the case when the stopping
time is large) random impulsive changes to the velocity of the
particle. FP found that the correlation time of the turbulent eddies
is typically $0.15$ orbits, which marginally satifies this
condition. Nonetheless we proceed with the model and later compare
with simulation results. In a time interval $(t_0,t)$, the change
$\delta v_{z}$ of the particle velocity produced by the random
acceleration $F$ can be written as
\begin{equation}\label{eq:fp3}
\delta
v_{z}=\int_{t_0}^{t}Fdt'=\int_{t_0}^{t}\frac{u_{z}(z,t')}{\tau_{\rm{st}}}dt'
\, .
\end{equation}
Because of the stochastic nature of $F$, this reduces to zero when a
suitable ensemble average is taken: $\langle \delta
v_{z}\rangle$=0. To describe the evolution of the particle
distribution, we need to calculate the mean square deviation of the
particle velocity:
\begin{equation}
\frac{1}{2}\left(\frac{d\,\left(\delta
v_{z}\right)^{2}}{dt}\right)=\int_{t_{0}}^{t}\frac{u_{z}(z,t)u_{z}(z,t')}{\tau_{\rm{st}}^{2}}dt'
\label{eq:fp4}
\end{equation}
Now, we take an ensemble average. When this is done, for an assumed
steady state and local homogeneous turbulence, the result of the
integral depends only on the time difference $t-t'$. Furthermore,
there is no prefered time so we finally get
\begin{equation}
\frac{d\,\langle \left(\delta
v_{z}\right)^{2}\rangle}{dt}=\frac{2}{\tau_{\rm{st}}^{2}}\int_{0}^{t}\langle
u_{z}(z,t)u_{z}(z,0)\rangle dt =
\frac{2}{\tau_{\rm{st}}^{2}}\int_{0}^{t}S_{zz}(t) dt.
\label{eq:fp7}
\end{equation}
Here $S_{zz}(t)$ is the velocity correlation function used in FP where
the reader is referred for more discussion of the above aspects.  At
large times, we thus obtain the limiting form
\begin{equation}
\frac{d\,\langle \left(\delta v_{z}\right)^{2}\rangle}{dt} \to
\frac{2D_{{\rm T}}(\infty)}{\tau_{\rm{st}}^{2}}
\label{eq:fp8}
\end{equation}
where $D_{\rm T}( \infty)\equiv \int_{0}^{\infty}\langle
u_{z}(z,\Delta t)u_{z}(z,0)\rangle d(\Delta t)$ defines an effective
diffusion coefficient and was already used by FP in a different
context in their discussion of small particles well coupled to the
gas, where it is the spatial diffusion coefficient.  Integrating the
last equation, we finally obtain for large time $t$
\begin{equation}
\langle (\delta v_{z})^{2}\rangle \sim \frac{2D_{\rm
T}(\infty)t}{\tau_{\rm{st}}^{2}}
\label{eq:fp10}
\end{equation}
The linear dependence in time in this expression shows that the
particles experience a random walk in velocity space, with an
effective diffusion coefficient
\begin{equation}
D_{\rm{v}}=\frac{D_{\rm T}(\infty)}{\tau_{\rm{st}}^{2}}.
\label{turb_coeff}
\end{equation}
We comment that if we set $t \sim \tau_{\rm {st}}$ in equation
(\ref{eq:fp10}), we expect that the mean square velocity dispersion
${\langle (\delta v_{z})^{2}}\rangle \sim 2D_{\rm
T}(\infty)/\tau_{\rm{st}},$ a form found by \citet{volketal80} from
considerations of stochastic forcing and a specific turbulence model.

\subsection{Fokker--Planck equation}

The next step is to determine the particle distribution function
$f=f(z,v_{z},t)$, from which quantities such as the vertical
semi-thickness can be determined. Its evolution in phase space can be
described by the one-dimensional Fokker-Planck equation. For the
system we study, the particles are subject to frictional drag and
gravity due to the central star as written in equation~(\ref{eq:fp2})
while their velocities diffuse in velocity space with diffusion
coefficient $D_{\rm{v}}$ given by equation~(\ref{eq:fp10}). Then, the
Fokker--Planck equation takes the form \citep{chandra49,johnsonetal06}:
\begin{equation}
\frac{\partial f}{\partial t}+v_{z}\frac{\partial f}{\partial
z}+\frac{\partial}{\partial v_{z}}\left( bf
\right)=D_{\rm{v}}\frac{\partial^{2}f}{\partial v_{z}^{2}} ,
\label{eq:fp11}
\end{equation}
where $b=-\Omega^2 z-v_{z}/\tau_{\rm{st}}$ is the non-stochastic
 acceleration.  The derivation of equation~(\ref{eq:fp11}) is given in
 Appendix~\ref{fp equation}.

We seek equilibrium solutions, expected to be attained on a time scale
comparable to the stopping time, for the distribution function found
by setting $\frac{\partial}{\partial t}=0$ in Eq. (\ref{eq:fp11}).  We
find a solution for $f$ of the form
\begin{equation}
f(z,v_z) \propto
e^{-\beta\left(\frac{v_{z}^{2}}{2}+\frac{\Omega^{2}z^{2}}{2}\right)},
\label{eq:fp16}
\end{equation}
where
\begin{equation}
\beta=\frac{\tau_{\rm{st}}}{D_{{\rm T}}(\infty)} .
\label{eq:fp16}
\end{equation}
Thus the thickness of the solid body layer $H_d$ is written as
\begin{equation}
H_d=\sqrt{\frac{D_{{\rm T}}(\infty)}{\tau_{\rm{st}}\Omega^{2}}}.
\label{eq:fp17}
\end{equation}
We note that this form of $H_d$ as the parameter of a Gaussian
distribution in $z$ is of the same from as applies to the strongly
coupled small particle case \citep[see][or FP]{dubrulleetal95}. Thus
the expression properly extends to all particle sizes.

\section{Numerical simulations}

In this section, we present a set of numerical simulations that
confirm the results of the discussion above. Those are local
simulations of stratified discs that use the shearing sheet
approximation \citep{hawleyetal95,stoneetal96}. The disc becomes
turbulent in the presence of a weak magnetic field. When turbulence is
well established, we follow the evolution of dust particles and larger
solid bodies using two different algorithms and compare the
steady--state thickness of the solid layer to the results presented
above. Below, we describe the set--up of the simulations before going
on to give our results.

\subsection{Method}

\begin{figure*}
\begin{center}
\includegraphics[scale=0.5]{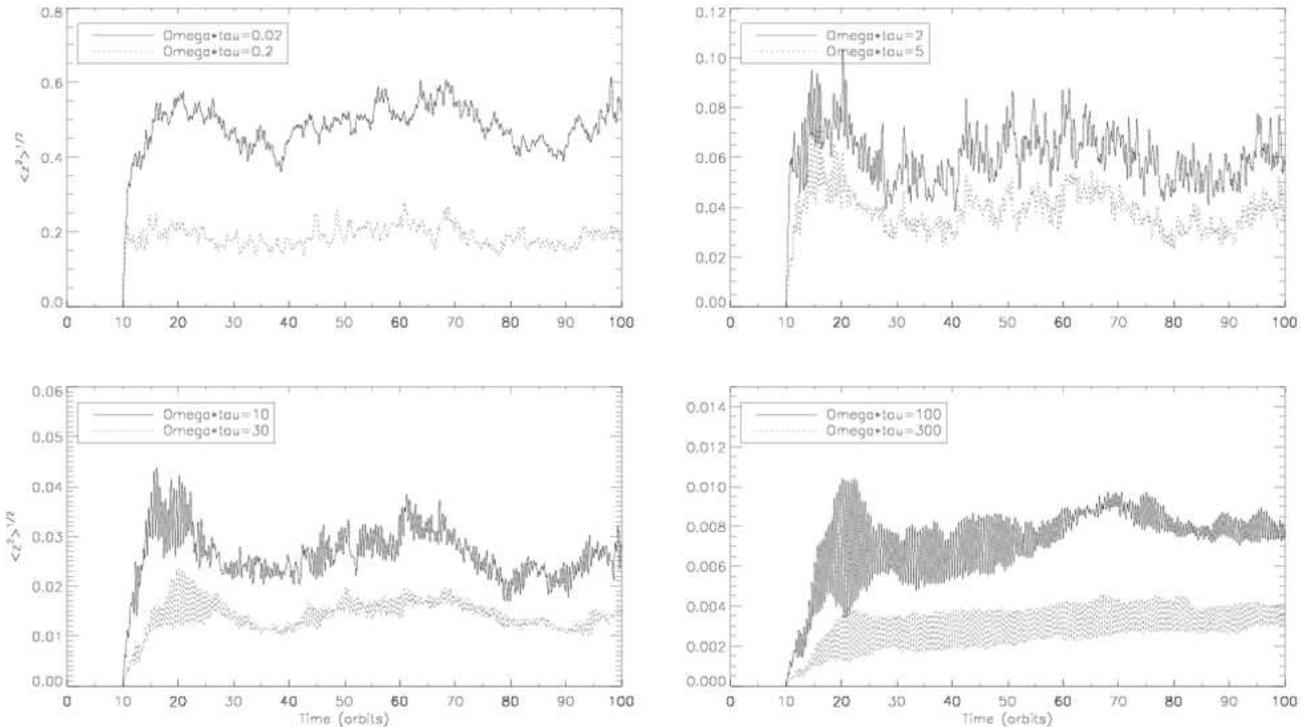}
\caption{Root mean square vertical displacement as a function of time
for the couplings $\Omega \tau_{\rm{st}}=0.02$, $0.2$, $2$, $5$, $10$,
$30$, $100$ and $300$ (given at the upper left hand corner of each
plot). Note the different scale used on each plot.}
\label{fig:rms_z_sq}
\end{center}
\end{figure*}

We used the ZEUS code \citep{stone&norman92a,stone&norman92b} to
evolve the standard MHD equations in a shearing box
\citep{goldreich&lyndenbell65}.  For the gas, we used a set--up
identical to that described in FP. In standard Cartesian coordinates
$(x,y,z)$, the box has a size $(H,2\pi H,6H)$ and we use a resolution
$(N_x,N_y,N_z)=(32,100,196)$.  We comment that the shearing box
represents a local patch of a differentially rotating disc and has
been used by many authors \citep[eg.][FP]{hawleyetal95,stoneetal96}
for the purposes of simulating MHD turbulence.  The coordinate origin
is at the centre of the box with the $x$ and $y$ axes pointing in the
outward radial direction and the direction of rotation respectively.
The length scale is $H,$ the time scale is set by $\Omega^{-1},$ and
mass scale is determined by the gas density scale. Here the
self--gravity of the gas is neglected. There is no
information concerning the distance to the centre of the disc other
than it should be much larger than $H.$ However, the shearing box can
be thought of as providing a detailed representation of a local region
of a disc for which $H$ and the distance to the centre are determined
by global considerations.
  
In our simulations, the equation of state is isothermal. The disc,
initially in hydrostatic equilibrium in the vertical direction, is
threaded by a magnetic field with zero net flux, such that the ratio
of thermal to volume averaged magnetic pressure, $\beta,$ is equal to
400. Because of the MRI, MHD turbulence develops and is maintained for
at least $100$ orbits. In agreement with previous published results
\citep{stoneetal96}, we found that $\alpha$, the ratio of the volume
averaged total stress (the sum of the $(x,y)$ components of the
Maxwell and Reynolds stress tensors) to $P_0,$ which governs the
angular nomentum transport, is typically $0.015$ throughout this type
of simulation (see FP for details).

This model was then used as a basis to investigate the evolution of an
ensemble of solids of different sizes. To do so, we used two different
algorithms. First, we adopted an N--body approach well suited to the
bodies of large size that we want to consider. In that case, the
velocities and positions of individual particles are updated in two
steps: the gas velocities are first interpolated at the position of
the particles using a bilinear interpolation method. Then, a second
order Runge--Kutta method is used to update the velocities and
positions of the particles in time according to their equation of
motion. For the smallest particles that we studied, we also used the
two--fluid algorithm presented by FP, in which the dust component is
modelled as a second pressureless fluid. In both cases, neither the
back reaction of the particles on the gas nor inter-particle
interactions are included.

The N--body approach was used to investigate the evolution of solid
bodies of $8$ different sizes. In each case, the particles were
assumed to be in the Epstein regime, so that $\tau_{\rm{st}}$ is
evaluated according to equation~(\ref{epstein_tau}). To label the
model, we used the value of the dimensionless parameter $\Omega
\tau_{\rm{st}}$, where $\tau_{\rm{st}}$ is evaluated in the midplane
of the unperturbed disc (i.e. the value of the density is that given
by the hydrostatic equilibrium). Following this convention, the $8$
models we ran correspond to $\Omega \tau_{\rm{st}}=0.02$, $0.2$, $2$,
$5$, $10$, $30$, $100$ and $300$. At $t=10$ orbits, we introduce $100$
particles for each of the models described above. They are distributed
in the midplane of the disc where they form a horizontal lattice
consisting of 25 columns and 4 rows of particles. Their initial
velocities are set equal to zero. The system is then evolved for a
further $90$ orbits. This long time integration enables a good
sampling of the flow properties for each particle. This is why a
larger number of particles is not required. Instead, we find that
time--averaging the results over the duration of the runs provides a
suitable statistical representation of the effect of MHD turbulence on
the particle distribution.

In addition, two runs were performed with the two--fluid method
mentioned above. They are limited to $\Omega \tau_{\rm{st}}=0.02$ and
$0.2$ (indeed, in order for the two--fluid approach to be valid, we
require $\Omega \tau_{\rm st} < 1$). As for the N--body calculation,
we assume that the Epstein law applies. The set--up we used for these
two--fluid models is identical to that of FP: at $t=20$ orbits, the
dust particles form a Gaussian thin disc such that $H_d=0.2 H$ and
their evolution is then followed for $80$ orbits.

\subsection{Results}
\label{num_results}

The results of the $8$ models that use the N--body approach are
illustrated in figure~\ref{fig:rms_z_sq}. For each of them, a curve is
plotted that shows the variations of the averaged root mean square
displacement $\langle Z^2 \rangle^{1/2}$ over all the particles with
time. The value of $\Omega \tau_{\rm{st}}$ corresponding to each case
is given in the upper left hand corner of each plot. For
convenience we adopt a system of units such that $H =1.$ However,
note the different scales on the vertical axes of the four panels. In
each model, $\langle Z^2 \rangle^{1/2}$ first increases before
oscillating aroung a mean value. This mean value represents the
average planetesimal disc semi-thickness. It is quickly well defined
for the smallest particles ($\Omega \tau_{\rm{st}} \leq 30$). For
$\Omega \tau_{\rm{st}}=100$ and $300$, more time is required before
reaching an equilibrium. This is because the interaction between the
particles and the gas is very weak in that case ($\tau_{\rm{st}}=15$
and $45$ orbits respectively).

\begin{figure}
\begin{center}
\includegraphics[scale=0.45]{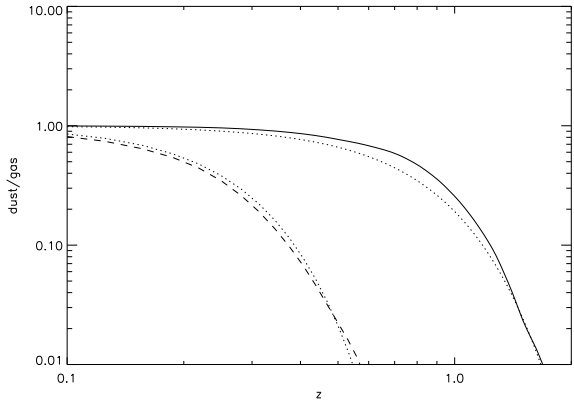}
\caption{Vertical profile for the dust to gas ratio in the two models,
  $\Omega \tau_{\rm{st}}=0.02$ ({\it solid line}) and $0.2$ ({\it
  dashed line}), in which the solid component is treated as a
  pressureless fluid. Both curves are time averaged between $t=65$ and
  $t=85$ orbits. The dotted line is a Gaussian fit to the data,
  using a thickness $H_d=0.55$ and $0.18$ respectively.}
\label{fig:two_fluid results}
\end{center}
\end{figure}

The results of the models that use the two--fluid approach are
illustrated in figure~\ref{fig:two_fluid results}. This shows the
vertical profile of the dust to gas mass density ratio, averaged in
time between $t=65$ and $t=85$ orbits for the two models:
$\Omega\tau_{\rm{st}}=0.02$ ({\it solid line}) and
$\Omega\tau_{\rm{st}}=0.2$ ({\it dashed line}). The {\it dotted}
curves show the function
\begin{equation}
\frac{\rho_d}{\rho} = \left( \frac{\rho_d}{\rho} \right)_0
\exp\left(-\frac{z^2}{2H_d^2} \right) \, ,
\end{equation}
with $H_d/H=0.55$ and $0.18$ respectively. These give the equilibrium
dust disc semi-thickness for these two models.

\begin{figure}
\begin{center}
\includegraphics[scale=0.45]{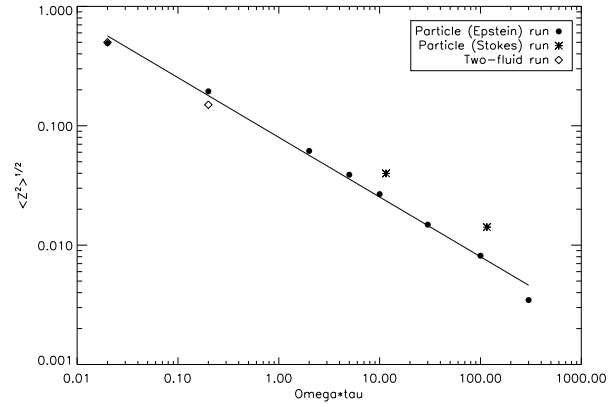}
\caption{Root-mean-square vertical displacement as a function of
dimensionless stopping time. The filled circles (Epstein drag) and the
asterisks (Stokes drag) represent data from simulations of individual
particles, while the two diamonds were obtained from calculations
regarding the dust component as a fluid. The straight line represents
the equation $\langle z^2 \rangle^{1/2}=0.08\,(\Omega
\tau_{\rm{st}})^{-1/2}$. }
\label{fig:rms_z}
\end{center}
\end{figure}

In Fig.~\ref{fig:rms_z}, we summarize the results of all models by
plotting the root mean square vertical extents as a function of
$\Omega \tau_{\rm{st}}$. The filled circles correspond to the
simulations performed with the N--body algorithm. They are obtained by
averaging the root mean square deviation of the particles between $50$
and $100$ orbits. The two diamonds are the data obtained from the
two-fluid calculations described above (the stars appearing in
figure~\ref{fig:rms_z} are further discussed in the discussion section
below). The straight line shows the equation
\begin{equation}
\langle z^{2} \rangle^{1/2}=0.08\,(\Omega \tau_{\rm{st}})^{-1/2} \, .
\label{eq:bestfit}
\end{equation}
which is the best fit to the data. It is seen to fit nicely the
results of the simulations. This is expected according to the
discussion given in section~\ref{model}. Indeed, using the relation
$H=c_s/\Omega$, equation~(\ref{eq:fp17}) can be expressed as
\begin{equation}
\frac{H_d}{H}=\sqrt{\frac{D_{\rm T}(\infty)}{c_s H}}(\Omega
\tau_{\rm{st}})^{-1/2} \, .
\end{equation}
From the correlation function of the turbulent velocities, FP found
$D_{\rm T}(\infty)=5.5 \times 10^{-3} c_s H$, which gives
\begin{equation}
\frac{H_d}{H}=7.4 \times 10^{-2}(\Omega \tau_{\rm{st}})^{-1/2} \, .
\label{th_results}
\end{equation}
This is in very good agreement with the results of the simulations
given by equation~(\ref{eq:bestfit}), so validating the discussion
presented in Section~\ref{model}.

\begin{figure}
\begin{center}
\includegraphics[scale=0.45]{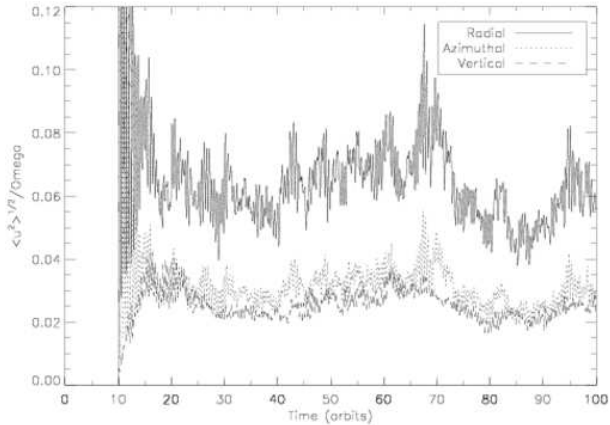}
\caption{The form of the root mean square velocity components
  divided by $\Omega$ as a function of time for the particles with
  $\Omega \tau_{\rm{st}}= 10.$ After initial transients, $\langle
  v_x^2\rangle^{1/2}\sim 2.5\langle v_z^2\rangle^{1/2}$ and $\langle
  v_y^2\rangle^{1/2}\sim 1.25\langle v_z^2\rangle^{1/2}.$ The $y$
  component is measured relative to the local mean shear.  }
\label{fig:rms_vel_sq}
\end{center}
\end{figure}

The evolution of $\langle v_x^2\rangle^{1/2},$ $\langle
v_y^2\rangle^{1/2}$ and $\langle v_z^2\rangle^{1/2}$, being the the
root mean square velocity components in the $x,y$ and $z$ directions,
respectively, as a function of time is illustrated in figure
\ref{fig:rms_vel_sq} for the particles with $\Omega \tau_{\rm{st}}=
10.$ Other values of the latter parameter give similar results. After
initial transients, noting that the $y$ component is measured relative
to the local mean shear, the ratios $\langle
v_y^2\rangle^{1/2}/\langle v_x^2\rangle^{1/2}\sim 0.5$
and $ \langle v_z^2\rangle^{1/2}/\langle v_x^2\rangle^{1/2}
\sim 0.4$ are established.  These can be understood as corresponding
to the root mean square turbulent gas velocity components which show
similar ratios \citep{hawleyetal95}.  In a shearing box simulation
which reaches a quasi steady state such ratios are established and
maintained. Restoring units, these velocity measures scale as $\Omega
H.$

\section{Discussion}

We have presented an analytical estimate of the thickness of the solid
layer formed by sedimenting solid bodies in a turbulent protoplanetary
disc. The stochastic dust vertical displacement in the presence of
turbulence is modelled using a Fokker-Planck equation, which describes
the evolution of the particle distribution function in phase
space. The steady state solution of this equation provides an
expression for the thickness $H_d$ of the disc of solids. We found
\begin{equation}
H_d/H \propto (\Omega \tau_{\rm{st}})^{-1/2} \, .
\end{equation}
To confirm this analytical result, we performed simulations of dust
particles embedded in a vertically stratified and turbulent local
model of a protoplanetary disc. We found the numerical values of the
particle height dispersions to agree well with the Fokker-Planck
result. These simulations were done utilising Epstein drag. In the
following, we discuss the expected differences that will arise when
Stokes drag is used.

\subsection{The Stokes drag law}

In the Stokes regime, the stopping time $\tau_{\rm{st}}$ is given by
equation~(\ref{stokes_tau}). For large particles, mostly confined in
the vicinity of the equatorial plane, we should be in the regime in
which $|\bb{v}| << |\bb{u}|$. In that case, an averaged expression for
the stopping time can be written as
\begin{equation}
\langle \tau_{\rm{st}} \rangle^{(St)} \sim \frac{8}{3C_D} \frac{\rho_s
  a}{\bar{\rho} c_s} \left( \frac{(\delta u_z^2)^{1/2}}{c_s}
  \right)^{-1} \, ,
\label{tau_stokes_avf}
\end{equation}
where $\bar{\rho}$ is the mean value of the turbulent density field
and $(\delta u_z^2)^{1/2}$ is the root mean square turbulent gas
vertical velocity fluctuation. We note that this expression gives a
very na\"ive estimate of the average value of $\langle \tau_{\rm{st}}
\rangle^{(St)}$. Indeed, $\tau_{\rm{st}}$, as defined in
equation~(\ref{stokes_tau}), depends on the turbulent velocity. As
such, the theory developed in section~\ref{diffusion vel} should be
modified and thus higher order velocity correlations would
appear. Such an analysis in detail is beyond the scope of this paper,
but the replacement of the quadratic velocity dependence in the drag
law by the product of a linear one and an average in the simple
estimate is likely to underestimate the particle force fluctuations
and accordingly underestimate the diffusion coefficient. As we will
see below, however, such effects remain small. In
equation~(\ref{tau_stokes_avf}), we have also considered $C_{\rm D}$
to be constant as in the large ${\cal R}_{e}$ limit. However, a
dependence of $C_{\rm D}$ on $(\delta u_z^2)^{1/2}$ could easily be
incorporated.

Using equation~(\ref{tau_stokes_avf}), the dimensionless parameter
$\langle \Omega \tau_{\rm{st}} \rangle^{(St)}$ can be written in terms
of the dimensionless parameter $\langle \Omega \tau_{\rm{st}}
\rangle^{(Ep)}$ corresponding to the Epstein regime, for which
\begin{equation}
\langle \Omega \tau_{\rm{st}} \rangle^{(Ep)}=\frac{\rho_s
    a}{\bar{\rho} c_s} \, .
\end{equation}
Indeed, we have:
\begin{equation}
\langle \Omega \tau_{\rm{st}} \rangle^{(St)}=\frac{8}{3C_D} \langle
  \Omega \tau_{\rm{st}} \rangle^{(Ep)} \left( \frac{(\delta
  u_z^2)^{1/2}}{c_s} \right)^{-1} \, .
\label{regime_relation}
\end{equation}
In our numerical simulations, we found $(\delta u_z^2)^{1/2}$ to be
 $\sim 0.13 c_s$ in the disc midplane. This is in agreement with
 previous results \citep{stoneetal96,turneretal06}. Furthermore, for
 the largest solid bodies, \citet{weidenschilling77} gives
 $C_D=0.44$. In that case, equation~(\ref{regime_relation}) gives
\begin{equation}
\langle \Omega \tau_{\rm{st}} \rangle^{(St)} \sim 45 \langle \Omega
  \tau_{\rm{st}} \rangle^{(Ep)}
\label{STOKE}
\end{equation}
This relation indicates that particles in the Stokes regime should
behave similarly to particles in the Epstein regime, but with a value
of the dimensionless parameter $\Omega \tau_{\rm{st}}$ that is $45$
times larger.

To check this, we performed two simulations in which the particles
follow the Stokes drag law, with $C_D=0.44$, and we measured the
equilibrium semi-thickness of the particle discs. The values of the
dimensionless parameter $\langle \Omega \tau_{\rm{st}} \rangle^{(Ep)}$
for these runs were taken to be equal to $0.24$ and $2.4$
respectively. According to the discussion above, the effective
dimensionless parameters $\langle \Omega \tau_{\rm{st}}
\rangle^{(St)}$ are respectively $\sim 11$ and $\sim 110$. The
measured particle disc thicknesses are represented in
figure~\ref{fig:rms_z} with stars.

 Although they lie close to the solid line (the vertical shift is
between $50 \%$ and a factor of 2), there seems to be a systematic
difference with the Epstein results. This arises because, as we
discussed above, the expression for the diffusion coefficient should
be adjusted upwards. Because of the quadratic dependence on velocities
of the drag force, we expect the points derived using the Stokes law
to lie systematically above the equivalent Epstein points, as is
observed in figure~\ref{fig:rms_z}. Taking the limitations of the
simple theory into account, the broad agreement of these simulations
with the Epstein results is quite satisfactory and confirms our
analysis.

\subsection{Gravitational instability}

Given the above results it is of interest to investigate whether
gravitational instability of a sedimenting layer in a protoplanetary
disc with the degree of turbulence considered here could be a
possibility. We begin by remarking that studies of the degree of
ionisation indicate that this may be adequate to allow the MRI to
operate both at small distances $\lesssim 0.2$ AU and large distances
$ \gtrsim 10-50$ AU from the central star \citep{fromang02}, depending
on the disc model. In the intermediate range, a dead zone is expected
to be present and affect the results presented below. We will
therefore concentrate on the outer parts of the disc in the following.

\citet[][\citeyear{weidenschilling95}]{weidenschilling94} has
 discussed the possible role of gravitational instability in the outer
 solar system in the formation of cometary aggregates. He considered
 bodies with $a \sim 10^{3-4}$ cm with a velocity dispersion
 determined by their radial drift driven by an underlying pressure
 gradient.  Whether such large bodies may be formed through the
 aggregation of small particles is
 uncertain. \citet{weidenschilling97} put forward a model applicable
 to a disc, without MHD turbulence, where bodies in this size range
 could aggregate by collisional sticking in the outer solar system on
 a $10^5$ year time scale. It is unclear to what extent the presence
 of MHD turbulence would extend this. The work described by
 \citet{weidenschilling&cuzzi93}, appropriate to the inner solar
 system, indicates that the time scale may be extended by about an
 order of magnitude, in which case they could form within expected
 lifetimes of protostellar discs.

Here we shall assume that bodies in the $10^4$ cm size range can form
and consider the effect of turbulence on the possibility of
gravitational instability in the outer protoplanetary disc. In the
first instance we neglect radial drift assuming that we are
considering an ensemble of solids near a pressure maximum and return
to this aspect later.

For the disc model, we adopt a surface density distribution as a
function of radius $r$ that follows that of the minimum mass solar
nebula \citep{hayashietal85}. The surface density of the solid disc,
$\sigma_p$, is given by
\begin{equation}
\sigma_p= \frac{\sigma_0}{\chi} \left(\frac{r}{5
\textrm{AU}}\right)^{-3/2} \, ,
\label{eq:sigma}
\end{equation}
where $\sigma_0$ is equal to 150 g.cm$^{-2}$ for the minimum mass
solar nebula and $\chi$ is the gas to dust ratio. The optical
depth of the disc of solids is
\begin{equation}
\tau=\frac{3\sigma_p}{4\rho_s a}.
\label{eq:optd}
\end{equation}
For the surface density profile (\ref{eq:sigma}) this gives
\begin{equation}
\tau=\frac{3\sigma_0}{4\rho_sa\chi}\left(\frac{r}{5
\textrm{AU}}\right)^{-3/2}.
\label{eq:sigmap}
\end{equation}
Thus for $r \ge 5$ AU, and $a \gtrsim 1$ cm, the disc of solid
material will be optically thin if we assume $\chi > 100\left(r/5
\textrm{AU}\right)^{-3/2}$.  It will then not be able to act
collectively as a fluid dust layer as in models that generate
turbulence through interaction with the gas
\citep[eg.][]{goldreich&ward73,weidenschilling80}.

Gravitational instability of the particle layer can occur if the
\citet{toomre64} criterion is satisfied:
\begin{equation}
\frac{c_{\rm{p}}\Omega}{\pi G \sigma_{\rm{p}}} < 1 \, ,
\label{eq:Toom}
\end{equation}
where $c_{\rm{p}}$ is the particle velocity dispersion in the
 radial or $x$ direction and $G$ is the gravitational constant. 
 The vertical scale height of the particle layer is
 $H_{\rm{d}}=\langle v_z^2 \rangle^{1/2}/\Omega$. Because the
 turbulent velocity fluctuations are larger in the radial direction
 than in the azimutal and vertical directions, the induced root mean
 square velocity component fluctuations induced in the solid bodies
 exhibit the same ordering (see figure~\ref{fig:rms_vel_sq} and
 above). We found $\langle v_x^2\rangle^{1/2}\sim 2.5 \langle
v_z^2\rangle^{1/2}$ and $\langle v_y^2\rangle^{1/2}\sim 1.25\langle
v_z^2\rangle^{1/2}.$ Thus $c_{\rm{p}} = \langle v_x^2\rangle^{1/2}
\sim 2.5 \langle v_z^2\rangle^{1/2}$. Equation~(\ref{eq:Toom}) can
then be expressed as
\begin{equation}
{H_d\over r} < \frac{3.5\times 10^{-6}}{\chi} \left(\frac{r}{5
  \textrm{AU}}\right)^{1/2} \left(\frac{\sigma_0}{1 \textrm{ g.cm}^{-2}}\right) \, ,
 \label{eq:Toom1}
\end{equation}
where the mass of the central star has been taken equal to one solar
mass and equation~(\ref{eq:sigma}) has been used. Using
Eq.~(\ref{th_results}), which gives the height of the dust layer
determined by turbulent diffusion, we find that
\begin{equation}
\Omega \tau_{\rm{st}} > 1.1\times 10^6 \chi^2 \left({H\over
  0.05r}\right)^2\left(\frac{r}{5
  \rm{AU}}\right)^{-1}\left(\frac{\sigma_0}{1 \textrm{ g.cm}^{-2}}\right)^{-2} \, .
\end{equation}
Assuming the particles are in the Epstein regime (at $50$ AU, this is
valid for particles smaller than approximately $1000$ metres) together
with $\rho_s=2$ g.cm$^{-3}$, a condition on their size can be found
from equation~(\ref{epstein_tau}) in the form
\begin{equation}
a > 2.8 \chi^2 \left({H\over 0.05r}\right)^2\left(\frac{r}{5
\textrm{AU}}\right)^{-5/2}\left(\frac{\sigma_0}{1 \textrm{ g.cm}^{-2}}\right)^{-1}\textrm{ km} \, .
\end{equation}
Thus at 50 AU, for $H/r =0.05$, gravitational instability becomes
possible for $ a > 600$ m for the minimum mass solar nebula if
$\chi=100$. But note that this quantity decreases with both $\sigma_0$
and $r$ and increases with $\chi$. For example, in a disc of only
$3$ times the minimum mass solar nebula, if $\chi=50$, solid boulders
larger than $\sim 50$ metres will become gravitationally unstable at
$50$ AU.

The length scale associated with the instability is $k^{-1} = H_d,$
with a characteristic mass $M_p \sim 6\pi \sigma_p H_{\rm{d}}^2$ which
gives
\begin{equation}
M_p \sim \frac{1.3\times 10^{18}}{\chi^3}\left(\frac{r}{5
\textrm{AU}}\right)^{3/2} \left( \frac{\sigma_0}{1 \textrm{ g.cm}^{-2}} \right)^3 \textrm{g} \, ,
\label{eq:Toom3}
\end{equation}
corresponding to a solid body of radius $\sim 5.4 (r/5
\textrm{AU})^{1/2} \chi^{-1} (\sigma_0/1$g.cm$^{-2}$) km. At $50$
AU, for the minimum mass solar nebula and when $\chi=100$, this
corresponds to a radius of $\sim 26$ km, similar to typical radii of
Kuiper belt objects \citep{lu&jewitt02}. In regions where
$\sigma_0/\chi$ or equivalently solid surface density is locally
enhanced, this size is accordingly increased. For example, for an an
enhancement of a factor of $10$, the masses involved in the
gravitational instability would correspond to bodies of $\sim 260$
km.
  
   As noted by \citet{weidenschilling95} this form of instability does
not mean that gravitational collapse ensues but that a gravitationally
bound cluster of total mass $M_p$ forms \citep[see
also][]{tangaetal04}. This then has to evolve through further settling
and accumulation. Nevertheless, the analysis presented here shows that
MHD turbulence does not necessarily prevent bodies of plausibly up to
few $100$ km in radius forming by gravitational instability in the
outer regions of protoplanetary discs.

Note that, in drawing this conclusion, we completely neglected the
effect of radial migration and concentrated only on vertical
settling. This is because, in general, the ratio of the radial drift
speed to orbital speed, induced by a radial pressure gradient, being
of order $(H/r)^2/(\Omega \tau_{\rm{st}})$ \citep[see][and references
therein]{pap&terquem06}, is significantly less than $H_d/r$ near
marginal stability for the conditions we consider, thus it can be
safely neglected.

\section*{ACKNOWLEDGMENTS}
We acknowledge the important contribution of James M. Stone towards
developping the N--body algorithm used in this paper. More details of
this will be described elsewhere. AC also wishes to acknowledge
support from CONACYT scholarship 167912. The simulations were
performed on the IoA cluster, University of Cambridge.

\bibliographystyle{mn2e} \bibliography{author}

\newcommand{\noopsort}[1]{}
\begin{thebibliography}{}

\bibitem[\protect\citeauthoryear{{Carballido}, {Stone} \&
  {Pringle}}{{Carballido} et~al.}{2005}]{carballidoetal05}
{Carballido} A.,  {Stone} J.~M.,    {Pringle} J.~E.,  2005, MNRAS, 358, 1055

\bibitem[\protect\citeauthoryear{{Chandrasekhar}}{{Chandrasekhar}}{1949}]{chan%
dra49}
{Chandrasekhar} S.,  1949, Reviews of Modern Physics, 21, 383

\bibitem[\protect\citeauthoryear{{Cuzzi}, {Dobrovolskis} \& {Champney}}{{Cuzzi}
  et~al.}{1993}]{cuzzietal93}
{Cuzzi} J.~N.,  {Dobrovolskis} A.~R.,    {Champney} J.~M.,  1993, Icarus, 106,
  102

\bibitem[\protect\citeauthoryear{{Dubrulle}, {Morfill} \& {Sterzik}}{{Dubrulle}
  et~al.}{1995}]{dubrulleetal95}
{Dubrulle} B.,  {Morfill} G.,    {Sterzik} M.,  1995, Icarus, 114, 237

\bibitem[\protect\citeauthoryear{{Fromang} \& {Papaloizou}}{{Fromang} \&
  {Papaloizou}}{2006}]{fromang&pap06}
{Fromang} S.,  {Papaloizou} J.,  2006, A\&A, 452, 751

\bibitem[\protect\citeauthoryear{{Fromang}, {Terquem} \& {Balbus}}{{Fromang}
  et~al.}{2002}]{fromang02}
{Fromang} S.,  {Terquem} C.,    {Balbus} S.~A.,  2002, MNRAS, 329, 18

\bibitem[\protect\citeauthoryear{{Goldreich} \& {Lynden-Bell}}{{Goldreich} \&
  {Lynden-Bell}}{1965}]{goldreich&lyndenbell65}
{Goldreich} P.,  {Lynden-Bell} D.,  1965, MNRAS, 130, 125

\bibitem[\protect\citeauthoryear{{Goldreich} \& {Ward}}{{Goldreich} \&
  {Ward}}{1973}]{goldreich&ward73}
{Goldreich} P.,  {Ward} W.~R.,  1973, ApJ, 183, 1051

\bibitem[\protect\citeauthoryear{{G{\'o}mez} \& {Ostriker}}{{G{\'o}mez} \&
  {Ostriker}}{2005}]{gomez&ostriker05}
{G{\'o}mez} G.~C.,  {Ostriker} E.~C.,  2005, ApJ, 630, 1093

\bibitem[\protect\citeauthoryear{{Hawley}, {Gammie} \& {Balbus}}{{Hawley}
  et~al.}{1995}]{hawleyetal95}
{Hawley} J.~F.,  {Gammie} C.~F.,    {Balbus} S.~A.,  1995, ApJ, 440, 742

\bibitem[\protect\citeauthoryear{{Hayashi}, {Nakazawa} \& {Nakagawa}}{{Hayashi}
  et~al.}{1985}]{hayashietal85}
{Hayashi} C.,  {Nakazawa} K.,    {Nakagawa} Y.,  1985, in {Black} D.~C.,
  {Matthews} M.~S.,  eds, Protostars and Planets II {Formation of the solar
  system}.
pp 1100--1153

\bibitem[\protect\citeauthoryear{{Johansen} \& {Klahr}}{{Johansen} \&
  {Klahr}}{2005}]{johansen&klahr05}
{Johansen} A.,  {Klahr} H.,  2005, ApJ, 634, 1353

\bibitem[\protect\citeauthoryear{{Johansen}, {Klahr} \& {Mee}}{{Johansen}
  et~al.}{2006}]{johansenetal06}
{Johansen} A.,  {Klahr} H.,    {Mee} A.,  2006, astroph/0603765

\bibitem[\protect\citeauthoryear{{Johnson}, {Goodman} \& {Menou}}{{Johnson}
  et~al.}{2006}]{johnsonetal06}
{Johnson} E.~T.,  {Goodman} J.,    {Menou} K.,  2006, ApJ, 647, 1413

\bibitem[\protect\citeauthoryear{{Luu} \& {Jewitt}}{{Luu} \&
  {Jewitt}}{2002}]{lu&jewitt02}
{Luu} J.~X.,  {Jewitt} D.~C.,  2002, ARAA, 40, 63

\bibitem[\protect\citeauthoryear{Natta, Testi, Calvet, Henning, Waters \&
  Wilner}{Natta et~al.}{2006}]{nattaetal06}
Natta A.,  Testi L.,  Calvet N.,  Henning T.,  Waters R.,    Wilner D.,  2006,
  astroph/0602041

\bibitem[\protect\citeauthoryear{{Papaloizou} \& {Terquem}}{{Papaloizou} \&
  {Terquem}}{2006}]{pap&terquem06}
{Papaloizou} J.~C.~B.,  {Terquem} C.,  2006, Rept.Prog.Phys., 69, 119

\bibitem[\protect\citeauthoryear{{Rodmann}, {Henning}, {Chandler}, {Mundy} \&
  {Wilner}}{{Rodmann} et~al.}{2006}]{rodmannetal06}
{Rodmann} J.,  {Henning} T.,  {Chandler} C.~J.,  {Mundy} L.~G.,    {Wilner}
  D.~J.,  2006, A\&A, 446, 211

\bibitem[\protect\citeauthoryear{{Stone}, {Hawley}, {Gammie} \&
  {Balbus}}{{Stone} et~al.}{1996}]{stoneetal96}
{Stone} J.~M.,  {Hawley} J.~F.,  {Gammie} C.~F.,    {Balbus} S.~A.,  1996, ApJ,
  463, 656

\bibitem[\protect\citeauthoryear{{Stone} \& {Norman}}{{Stone} \&
  {Norman}}{1992a}]{stone&norman92a}
{Stone} J.~M.,  {Norman} M.~L.,  1992a, ApJS, 80, 753

\bibitem[\protect\citeauthoryear{{Stone} \& {Norman}}{{Stone} \&
  {Norman}}{1992b}]{stone&norman92b}
{Stone} J.~M.,  {Norman} M.~L.,  1992b, ApJS, 80, 791

\bibitem[\protect\citeauthoryear{{Tanga}, {Weidenschilling}, {Michel} \&
  {Richardson}}{{Tanga} et~al.}{2004}]{tangaetal04}
{Tanga} P.,  {Weidenschilling} S.~J.,  {Michel} P.,    {Richardson} D.~C.,
  2004, A\&A, 427, 1105

\bibitem[\protect\citeauthoryear{Toomre}{Toomre}{1964}]{toomre64}
Toomre A.,  1964, ApJ, 139, 1217

\bibitem[\protect\citeauthoryear{{Turner}, {Willacy}, {Bryden} \&
  {Yorke}}{{Turner} et~al.}{2006}]{turneretal06}
{Turner} N.~J.,  {Willacy} K.,  {Bryden} G.,    {Yorke} H.~W.,  2006, ApJ, 639,
  1218

\bibitem[\protect\citeauthoryear{{Voelk}, {Morfill}, {Roeser} \&
  {Jones}}{{Voelk} et~al.}{1980}]{volketal80}
{Voelk} H.~J.,  {Morfill} G.~E.,  {Roeser} S.,    {Jones} F.~C.,  1980, A\&A,
  85, 316

\bibitem[\protect\citeauthoryear{{Weidenschilling}}{{Weidenschilling}}{1977}]{%
weidenschilling77}
{Weidenschilling} S.~J.,  1977, MNRAS, 180, 57

\bibitem[\protect\citeauthoryear{{Weidenschilling}}{{Weidenschilling}}{1980}]{%
weidenschilling80}
{Weidenschilling} S.~J.,  1980, Icarus, 44, 172

\bibitem[\protect\citeauthoryear{{Weidenschilling}}{{Weidenschilling}}{1994}]{%
weidenschilling94}
{Weidenschilling} S.~J.,  1994, Nature, 368, 721

\bibitem[\protect\citeauthoryear{{Weidenschilling}}{{Weidenschilling}}{1995}]{%
weidenschilling95}
{Weidenschilling} S.~J.,  1995, Icarus, 116, 433

\bibitem[\protect\citeauthoryear{{Weidenschilling}}{{Weidenschilling}}{1997}]{%
weidenschilling97}
{Weidenschilling} S.~J.,  1997, Icarus, 127, 290

\bibitem[\protect\citeauthoryear{{Weidenschilling} \& Cuzzi}{{Weidenschilling}
  \& Cuzzi}{1993}]{weidenschilling&cuzzi93}
{Weidenschilling} S.~J.,  Cuzzi J.~N.,  1993, in Protostars and Planets III
  {Formation of lanetesimals in the solar nebula}.
pp 1031--1060

\bibitem[\protect\citeauthoryear{{Youdin} \& {Shu}}{{Youdin} \&
  {Shu}}{2002}]{youdin&shu02}
{Youdin} A.~N.,  {Shu} F.~H.,  2002, ApJ, 580, 494

\end{thebibliography}

\appendix

\section{The Fokker-Planck equation}
\label{fp equation}

The purpose of this appendix is to present a derivation of the
Fokker--Planck equation~(\ref{eq:fp11}) that we used in
section~\ref{model} to estimate the planetesimal disc thickness. This
derivation is largely based on \citet{chandra49} in which applications
of the theory of Brownian motion to dynamical friction and stellar
dynamics are discussed.

Consider the motion of a given solid body in the vertical direction.
During a small time $\Delta t$ , it will undergo a change in velocity
$\Delta v_z$. Some of this change is due to stochastic forces
that arise from the turbulent gas, and which have correlation times
that are assumed to be smaller than the time required for typical
velocities of the solid bodies to change significantly.
Also velocity changes induced during a characteristic correlation time
are assumed to be small. This is well satisfied for the large
particles we consider in this paper. From equation~(\ref{eq:fp2}), we
know that $\Delta v_z$ relates to the stochastic forcing term $F$
through
\begin{equation}
F(\Delta t)=\Delta v_z + \Omega^2 z \Delta t + (v_z/\tau )\Delta t
\label{motion_eq}
\end{equation}
The probability distribution function of $F(\Delta t)$ is known from
the theory of Brownian motion:
\begin{equation}
\Psi (z,v_z,\Delta v_z,\Delta t)=\frac{1}{\sqrt{4\pi q \Delta t}} \exp
\left( - \frac{F(\Delta t)^2}{4 q \Delta t} \right) \, ,
\label{psi_function}
\end{equation}
where $F(\Delta t)$ is given by equation~(\ref{motion_eq}) and $q
\equiv D_v$ is the diffusion coefficient associated with the diffusive
evolution of the velocity. We demonstrated in section~{\ref{diffusion
vel}} that it relates to the velocity correlation function through
equation~(\ref{turb_coeff}).

Consider now the ensemble constituted by all the particles in the
volume of interest. Let $f(z,v_z,t)$ be its distribution
function. Under the assumption that the turbulence is a
stationary stochastic process with finite amplitude and correlation
time \citep{johnsonetal06}, its time evolution relates to the
probability function $\Psi$ through:
\begin{eqnarray}
f(z,v_z,t+\Delta t)=\int_{-\infty}^{+\infty} f(z-v_z \Delta
t,v_z-\Delta v_z,t) \times \nonumber \\ \Psi(z-v_z \Delta t,v_z-\Delta
v_z,\Delta v_z,\Delta t) d(\Delta v_z)
\end{eqnarray}
This equation can be expanded to first order in $\Delta t$ to give
\begin{equation}
\frac{\partial f}{\partial t} + v_z \frac{\partial f}{\partial z} +
  \frac{\partial }{\partial v_z} \left[ \frac{(\Delta
  v_z)_{av}}{\Delta t} f \right] = \frac{1}{2} \frac{\partial^2
  }{\partial v_z^2} \left[ \frac{(\Delta v_z^2)_{av}}{\Delta t} f
  \right]
\label{fp1}
\end{equation}
where the quantities $(\Delta v_z)_{av}$ and $(\Delta v_z^2)_{av}$ are
defined by the relations:
\begin{eqnarray}
(\Delta v_z)_{av} &=& \int_{-\infty}^{+\infty} \Delta v_z
  \Psi(z,v_z,\Delta v_z,\Delta t) d(\Delta v_z) \, , \\ (\Delta
  v_z^2)_{av} &=& \int_{-\infty}^{+\infty} (\Delta v_z)^2
  \Psi(z,v_z,\Delta v_z,\Delta t) d(\Delta v_z) \, .
\end{eqnarray}
Using equation~(\ref{psi_function}), both quantities can be evaluated
to first order in $\Delta t$:
\begin{eqnarray}
(\Delta v_z)_{av} &=& (-\frac{v_z}{\tau_{\rm{st}}}-\Omega^2 z) \Delta
  t \, , \label{av} \\ (\Delta v_z^2)_{av} &=& 2 q \Delta t
  \label{av2} \, .
\end{eqnarray}
After eliminating $\Delta t$, we finally obtain
equation~(\ref{eq:fp11}) by substituting equations~(\ref{av}) and
(\ref{av2}) in equation~(\ref{fp1}).

\end{document}